\documentclass[aps,prl,reprint,superscriptaddress]{revtex4-1}

\usepackage{amsmath}
\usepackage{amssymb}
\usepackage{enumitem}
\usepackage{times}

\usepackage[hiresbb,dvipdfmx]{graphicx}

\usepackage{xcolor}
\usepackage{lipsum}

\newcommand{\Star}[1]{#1\ensuremath{^*}\kern-\scriptspace}
\newcommand{\CStar}{\Star{\ensuremath{\mathrm{C}}}}

\newcommand{\prlsection}[1]{{\par\it #1.---}}

\begin{document}

%\preprint{}

\title{Geometric Formulation of Universally Valid Uncertainty Relation for Error}


\author{Jaeha Lee}
\email[]{lee@iis.u-tokyo.ac.jp}
%\thanks{}
\affiliation{Institute of Industrial Science, The University of Tokyo, Chiba 277-8574, Japan.}

\author{Izumi Tsutsui}
\email[]{izumi.tsutsui@kek.jp}
\affiliation{Theory Center, Institute of Particle and Nuclear Studies, High Energy Accelerator Research Organization (KEK), Ibaraki 305-0801, Japan.}


%\date{\today}

\begin{abstract}
We present a new geometric formulation of uncertainty relation valid for any quantum measurements of statistical nature. Owing to its simplicity and tangibility, our relation is universally valid and experimentally viable. Although our relation violates the na{\"i}ve non-commutativity bound $\hbar/2$ for the measurement of position and momentum, the spirit of the uncertainty principle still stands strong. Our relation entails, among others, the Ozawa relation as a corollary, and also reduces seamlessly to the standard Kennard-Robertson relation when the measurement is non-informative.
\end{abstract}

%\keywords{}

%\maketitle must follow title, authors, abstract,  and keywords
\maketitle


%%%%%
%%%%%%%%%%
\prlsection{Introduction\label{sec:intro}}
%%%%%%%%%%
%%%%%
The uncertainty principle stands undoubtedly as one of the basic tenets of quantum mechanics, characterizing the indeterministic nature of the microscopic world.  Soon after Heisenberg's seminal exposition \cite{Heisenberg_1927}, the first mathematical formulation of the uncertainty principle was presented by Kennard \cite{Kennard_1927} giving the lower bound $\hbar/2$ for the product of the standard deviations of position and momentum, which was later generalized to those of arbitrary observables by Robertson \cite{Robertson_1929}.  Owing to its mathematical clarity and simplicity, the Kennard-Robertson (KR) inequality became a standard textbook material as a succinct expression of quantum indeterminacy, and has since been regarded widely as {\it the} uncertainty relation in the general discourse, despite the fact that it has little to do with measurement.

Meanwhile, even though his own conception of uncertainty (or \lq indeterminateness\rq\ \cite{Heisenberg_1930}) is difficult to precisely infer from the rather vague description of his writings, Heisenberg did entertain concepts of error and disturbance associated with measurement when he mentioned various examples such as the famous gamma ray microscope experiment.  This somewhat unsatisfying status led to the emergence of several alternative formulations of uncertainty relations involving measurement, typically adopting the indirect measurement scheme where the system of the measuring device is considered explicitly in addition to the target system of one's interest.  There, an operational viewpoint was incorporated into the concepts of error and disturbance, which resulted in, {\it e.g.}, the Arthurs-Kelly-Goodman (AKG) inequalities \cite{Arthurs_1965,Arthurs_1988} and the more recent Ozawa inequalities \cite{Ozawa_2003, Ozawa_2004_01} along with their refinements \cite{Branciard_2013}.  Apart form these, uncertainty relations have also been analyzed from a measure-theoretic viewpoint \cite{Werner_2004} as well as within the framework of estimation theory \cite{Watanabe_2011}.

Beyond the orthodox relations regarding error and disturbance, the uncertainty principle has also been found to lie at the heart of many other intriguing physical phenomena, leading to the discovery of various types of trade-off relations regarding, {\it e.g.},  time and energy \cite{Mandelshtam_1945, Allcock_1969_01, Allcock_1969_02, Allcock_1969_03, Helstrom_1976}, entropy \cite{Hirschman_1957, Beckner_1975, Birula_1975, Deutsch_1983}, conservation law \cite{Wigner_1952, Araki_1960, Yanase_1961, Ozawa_2002_01}, speed limit \cite{Fleming_1973, Aharonov_1990, Pfeifer_1993, Margolus_1998, Giovannetti_2003, Jones_2010, Pires_2016, Shiraishi_2018}, gate implementation \cite{Ozawa_2002_02, Tajima_2018}, and counterfactuality \cite{Hall_2001, Dressel_2014, Lee_2016, Pollak_2019}.

In this Letter, we present a novel uncertainty relation that marks the trade-off relation between the measurement errors of two arbitrary quantum observables.  We note three distinctive merits that characterize our formulation.  
First, it is established upon a conceivably simplest framework of measurement without any reference to the specific measurement models whatsoever: the only objects we deal with are the tangible measurement outcomes.  This implies that our relation is universally valid as well as operationally useful, and is free from the problems some alternative formulations are know for, in which the error (or the disturbance) is to be evaluated from a set of observables which may not be measurable simultaneously \cite{Werner_2004, Koshino_2005}.  
Second, despite the fact that our uncertainty relation violates generically the non-commutativity bound,  which is in line with the recent similar findings espoused notably by Ozawa \cite{Ozawa_2004_01}, the spirit of the uncertainty principle still stands strong as a more general qualitative statement than is commonly conceived with clear physical and statistical meanings for its lower bound.
Third, our geometric formulation is capable of expressing various types of trade-off relations of different nature within a unified framework, thereby providing a seamless connection among the various forms in which the uncertainty principle manifests itself.  
In other words, our uncertainty relation acts as a \lq master relation\rq\ from which various known uncertainty relations can be derived, including the KR, AKG and Ozawa inequalities mentioned above.  Apart from the derivation of the KR inequality and the outline leading to the AKG and one of the Ozawa inequalities, we shall report the details on the physical ramifications of our geometric framework and its mathematical description in our subsequent papers.

%%%%%
%%%%%%%%%%
\prlsection{Quantum Measurement\label{sec:setup}}
%%%%%%%%%%
%%%%%
Let us first present our geometric framework.  We start by introducing the state space of a quantum system modeled as the convex set $Z(\mathcal{H})$ of all the density operators $\rho$ on a Hilbert space $\mathcal{H}$.  Its classical counterpart is the convex set $W(\Omega)$ of all the probability distributions $p$ on a sample space $\Omega$.  Our primary objects of investigation are affine maps $M : Z(\mathcal{H}) \to W(\Omega)$ from quantum state spaces to classical state spaces, {\it i.e.}, maps that take density operators $\rho$ to probability distributions $p = M\!\rho$ while maintaining the structure of the probabilistic mixture  $M(\lambda \rho_{1} + (1-\lambda) 
_{2}) = \lambda M\!\rho_{1} + (1-\lambda) M\!\rho_{2}$ for $\rho_{1}, \rho_{2} \in Z(\mathcal{H})$, $0 \leq \lambda \leq 1$.

%%%%%
\begin{figure}
\includegraphics[hiresbb,clip,height=7cm]{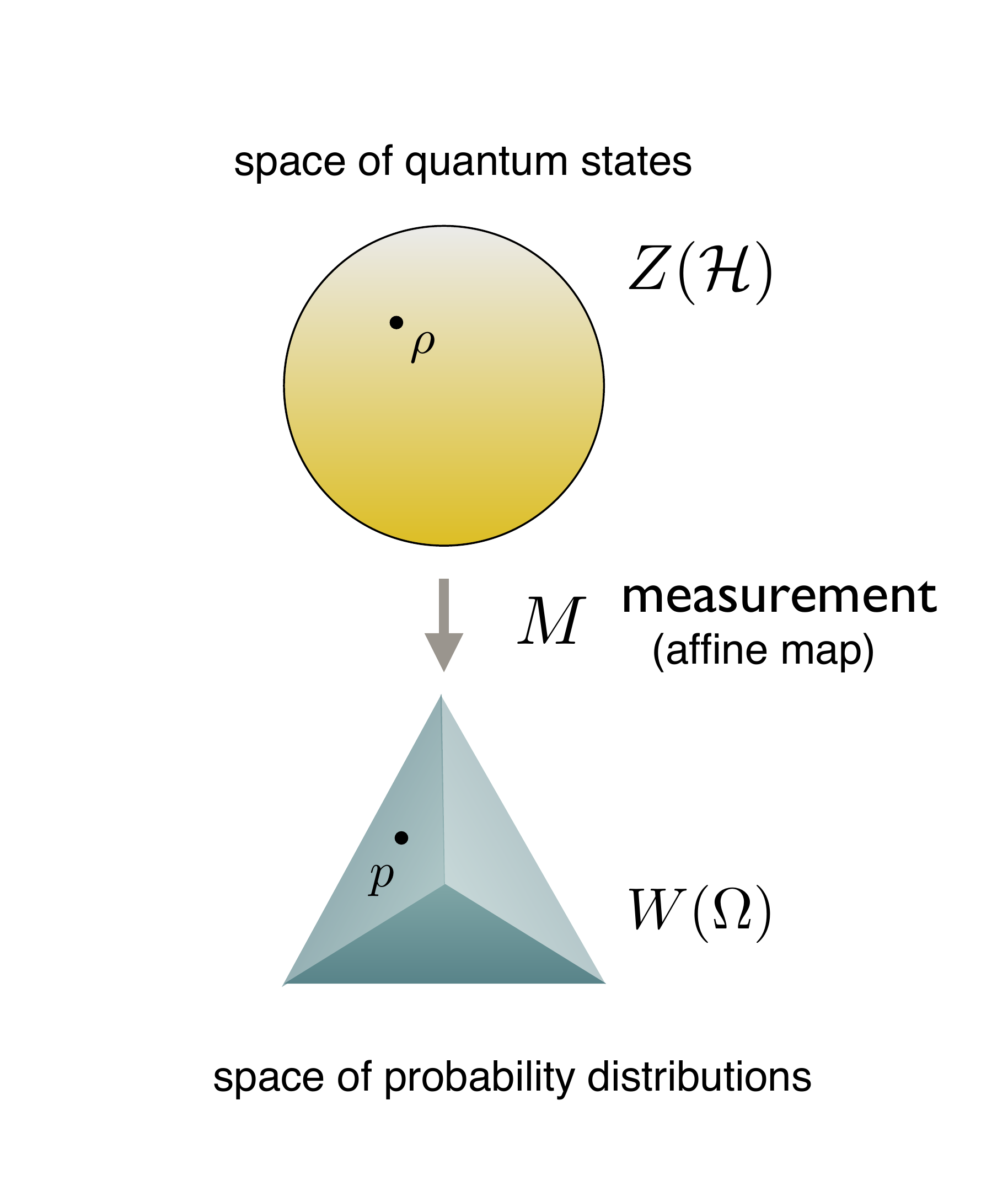}%
\caption{Our basic premise of quantum measurements.  The space of quantum states (density operators) $Z(\mathcal{H})$ is depicted as a sphere, while the space of probability distributions $W(\Omega)$ is depicted as a tetrahedron.  Quantum measurement $M$ can in general be regarded as a map $M : Z(\mathcal{H}) \to W(\Omega)$, given the fact that in any measurement (of statistical nature) all we have at the end is a probability distribution $p \in W(\Omega)$ over the possible set of outcomes associated with the quantum state $\rho \in Z(\mathcal{H})$ of our concern.  
\label{fig:one}}
\end{figure}
%%%%%

The map $M$ generally admits a wide range of interpretations, such as the representation of a quantum system by a classical model, but for the purpose of this Letter, let it be understood as a quantum measurement.  It is not difficult to convince oneself that this interpretation is indeed possible by considering the archetypal projection measurement associated with an arbitrary quantum observable $\hat{M}$.  
More explicitly, the spectral decomposition $\hat{M} = \sum_{i=1}^{N} m_{i}\, |m_{i}\rangle\langle m_{i}|$ of an observable on an $N$-dimensional Hilbert space induces a natural map
\begin{equation}\label{def:projection_measurement}
M : \rho \mapsto (M\!\rho)(m_{i}) := \mathrm{Tr}[ |m_{i}\rangle\langle m_{i}| \rho ]
\end{equation}
defined by the Born rule.  It is easy to see that \eqref{def:projection_measurement} is an affine map that takes a density operator $\rho$ to a probability distribution $M\!\rho$ on the sample space $\Omega := \{ m_{1}, \dots, m_{N} \}$ consisting of the observable's eigenvalues.  These eigenvalues are regarded as the possible measurement outcomes, and the probability distribution given in \eqref{def:projection_measurement} provides the probability $p(m_{i}) = (M\!\rho)(m_{i})$ of finding the respective outcomes $m_{i}$ in the measurement, which we now identify with the map $M$ as a whole (see FIG. \ref{fig:one}).  Throughout this Letter, the reader may safely assume the map $M$ to be that of the familiar projection measurement described above without missing much of the essence of the subject, although our $M$ is by no means restricted to that particular class.  In fact, the sole constraint we impose on $M$, {\it i.e.}, affineness, is indispensable for the self-consistent statistical interpretation of density operators: the outcome probability distribution should be invariant under every (pure-state) decomposition of a mixed quantum state.  In other words, our $M$ effectively belongs to the broadest class of quantum measurements ever conceivable of statistical nature.

An important observation is that a quantum measurement $M$ uniquely induces a map $M^{\prime}$ that takes functions on $\Omega$ to operators on $\mathcal{H}$.  This dual notion of a quantum measurement, termed its adjoint, is uniquely characterized by the relation
\begin{equation}\label{char:adjoint_quantum-measurement}
\langle M^{\prime}f \rangle_{\rho} = \langle f \rangle_{M\!\rho}
\end{equation}
valid for all complex functions $f$ on $\Omega$ and quantum states $\rho$ on $\mathcal{H}$.  Here, we have introduced the shorthand $\langle X \rangle_{\!\rho} := \mathrm{Tr}[X\rho]$ for a given pair of a Hilbert space operator $X$ and a density operator $\rho$ on $\mathcal{H}$, as well as $\langle f \rangle_{p} := \int_{\Omega} f(\omega) p(\omega)\, d\omega$ for a pair of a complex function $f$ and a probability distribution $p$ on $\Omega$.  Again, the projection measurement \eqref{def:projection_measurement} provides a prime example, the adjoint of which can be confirmed to read
\begin{equation}\label{def:adjoint_projection_measurement}
M^{\prime}f := \sum_{i=1}^{N} f(m_{i})\, |m_{i}\rangle\langle m_{i}|
\end{equation}
which fulfills \eqref{char:adjoint_quantum-measurement}.  Projection measurements are convenient in that they admit concrete expressions for the measurement \eqref{def:projection_measurement} and its adjoint \eqref{def:adjoint_projection_measurement} using familiar notions, allowing for the verification of the various claims in this Letter by means of direct computation.  (A rigorous proof for general affine $M$ will be given elsewhere \cite{Lee_2020_Math}.) 

\begin{figure}
\includegraphics[hiresbb,clip,height=6.3cm]{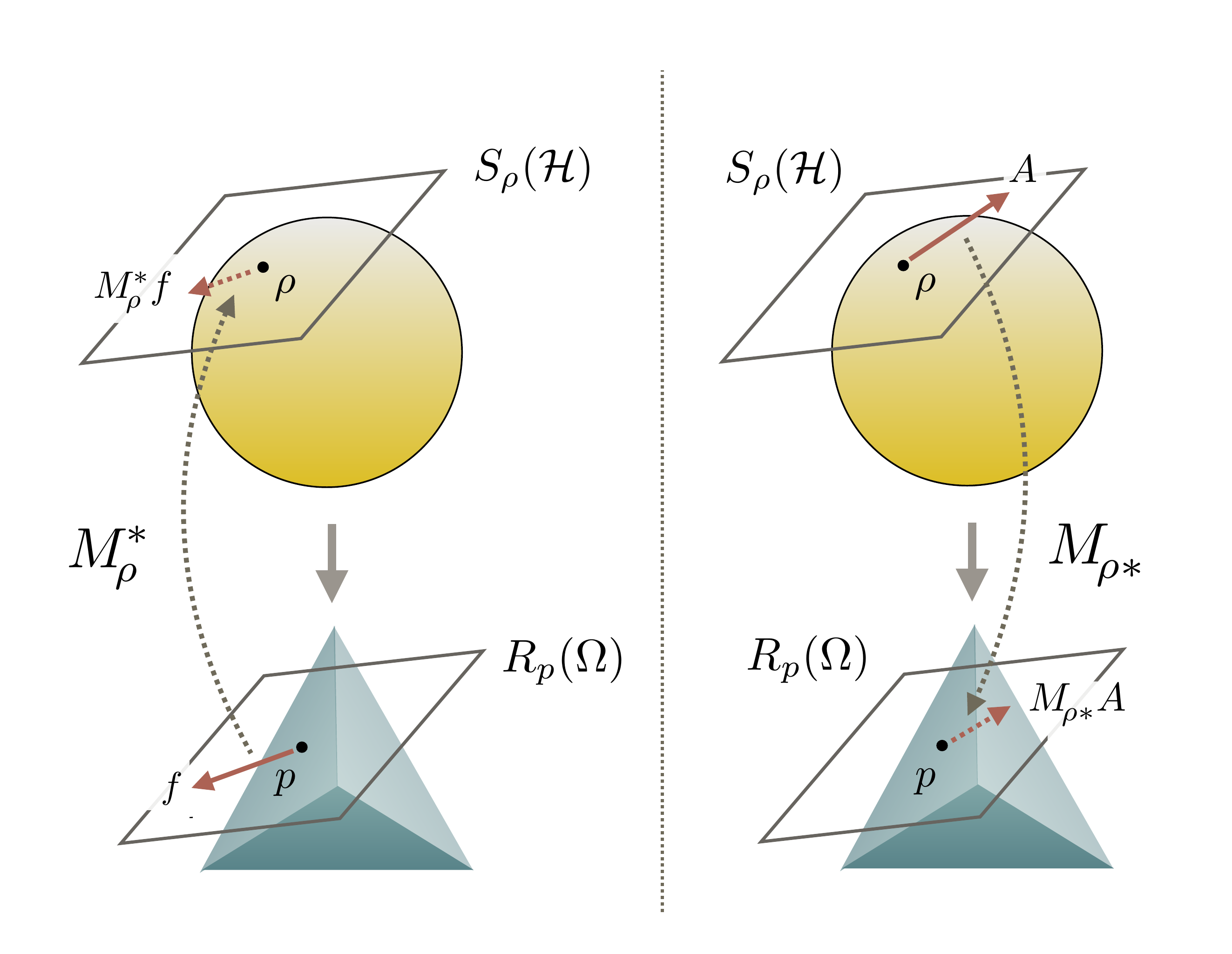}%
\caption{The pullback and the pushforward of the measurement.  (Left) Our quantum measurement map $M$ entails the pullback $M_{\!\rho}^{\hspace{-0.5pt}\ast}$ from the space of 
real functions $R_{p}(\Omega)$ attached to each of the points $p = M\!\rho \in W(\Omega)$ to the space of self-adjoint operators $S_{\rho}(\mathcal{H})$ attached to each of the points $\rho \in Z(\mathcal{H})$.  (Right) Conversely, the inner products furnished on each of the spaces $R_{p}(\Omega)$ and $S_{\rho}(\mathcal{H})$ uniquely induces the pushforward 
$M_{\!\rho\ast}{\hspace{-0.5pt}}$
as the adjoint to the pullback.  Both the pullback and the pushforward are contractions, that is, the norm decreases (or remains unchanged) under each of the maps.
\label{fig:pullback_pushforward}}
\end{figure}

%%%%%
%%%%%%%%%%
\prlsection{Pushforward and Pullback\label{sec:pushforward_pullback}}
%%%%%%%%%%
%%%%%
The key element of our framework is the fact that a quantum measurement, which is a global map between state spaces, gives rise to an adjoint pair of {\it local} ({\it i.e.}, state-dependent) maps which allow us to evaluate the accuracy of the measurement $M$ of an observable $A$ with respect to a function $f$ (see FIG. \ref{fig:pullback_pushforward}).
To expound on this, let us introduce the observable space of a quantum system modeled as the linear space $S(\mathcal{H})$ of all the self-adjoint ({\it alias} Hermitian) operators on $\mathcal{H}$.  Given a self-adjoint operator $A$, each quantum state $\rho \in Z(\mathcal{H})$ furnishes a seminorm $\lVert A \rVert_{\rho} := \sqrt{\langle A^{\dagger} A \rangle_{\!\rho}}$ on $S(\mathcal{H})$ that allows for the identification $A \sim_{\rho} B \iff \lVert A - B \rVert_{\rho} = 0$ of quantum observables into their equivalence classes.  These equivalence classes collectively form a quotient space, the completion of which we denote by $S_{\rho}(\mathcal{H})$.  By a similar procedure, a probability distribution $p \in W(\Omega)$ induces a seminorm $\lVert f \rVert_{p} := \sqrt{\langle f^{\dagger}f \rangle_{p}}$ on the space $R(\Omega)$ of all the real functions on the sample space $\Omega$.  The completion of the quotient space induced by the identification $f \sim_{p} g \iff \lVert f - g \rVert_{p} = 0$ on $R(\Omega)$ will be denoted by $R_{p}(\Omega)$.  As commonly practiced, we make a slight abuse of notation to denote the equivalence class with one of its representatives.  Also, in the above we have used the adjoint $A^{\dagger}$ and the complex conjugate $f^{\dagger}$ to expose the structure of the seminorm, although they are equivalent to $A$ and $f$, respectively, for $A \in S(\mathcal{H})$ and $f \in R(\Omega)$.

%%%%%
%%%%%

An important observation regarding quantum measurements $M : Z(\mathcal{H}) \to W(\Omega)$ is the validity of the inequality
\begin{equation}\label{ineq:contractivity}
\lVert f \rVert_{M\!\rho}^{\phantom{\ast}} \geq \lVert M^{\prime} f \rVert_{\rho}^{\phantom{\ast}}
\end{equation}
for any quantum state $\rho$ on $\mathcal{H}$ and complex function $f$ on $\Omega$, with the definitions of the respective seminorms being extended verbatim to those for normal operators and complex functions.  This can be understood as a corollary to the Kadison-Schwarz inequality \cite{Kadison_1952}, which is in a sense a generalization of the renowned Cauchy-Schwarz inequality to \CStar-algebras.  Indeed, its straightforward application to the adjoint $M^{\prime}$ yields the evaluation $M^{\prime}(f^{\dagger}f) \geq (M^{\prime}f)^{\dagger}(M^{\prime}f)$ valid for any complex function $f$.  This, combined with the characterization \eqref{char:adjoint_quantum-measurement} of the adjoint, leads to the desired result.

An intuitive operational interpretation of inequality \eqref{ineq:contractivity} is also available.  For this, first observe its equivalence to the condition $\sigma_{M\!\rho}(f) \geq \sigma_{\rho}(M^{\prime}f)$, where the respective symbols $\sigma_{\rho}(A) := \sqrt{\lVert A \rVert_{\rho}^{2} - \langle A \rangle_{\rho}^{2}}$ and $\sigma_{p}(f) := \sqrt{\lVert f \rVert_{p}^{2} - \langle f \rangle_{p}^{2}}$ denote the quantum and classical standard deviations.  This allows us to frame \eqref{ineq:contractivity} as a statement regarding the efficiency of the measurement: the operational cost of acquiring the expectation value of an observable through measurements can never overcome the quantum limit imposed by the said observable.  To illustrate this, consider obtaining the expectation value of an observable $A = M^{\prime}f$ by means of a measurement $M$ with respect to some real function $f$, which we call an estimator of the observable.  The inequality states that the operational cost $\sigma_{M\!\rho}(f)$ of the acquisition of the expectation value $\langle f \rangle_{M\!\rho} = \langle A \rangle_{\rho}$ is bounded from below by the quantum limit $\sigma_{\rho}(A)$ imposed by the observable.

An immediate consequence of inequality \eqref{ineq:contractivity} is the implication $f \sim_{M\!\rho} g \implies M^{\prime}f \sim_{\rho} M^{\prime}g$.  This allows the adjoint $M^{\prime}$, which was initially introduced as a map from functions to operators, to be passed to the map from equivalence classes of functions to that of operators.  We call the resultant map
\begin{equation}\label{def:pullback}
M_{\!\rho}^{\hspace{-0.5pt}\ast} : R_{M\!\rho}(\Omega) \to S_{\rho}(\mathcal{H})
\end{equation}
between quotient spaces the {\it pullback of the measurement} $M$ over the quantum state $\rho$.  
In concrete terms, this implies that, given (the equivalence class of) a real function $f \in R_{M\!\rho}(\Omega)$, we have (that of) a corresponding self-adjoint operator $M_{\!\rho}^{\hspace{-0.5pt}\ast} f \in S_{\rho}(\mathcal{H})$.
Note that, by construction, we have $\lVert f \rVert_{M\!\rho} \geq \lVert M_{\!\rho}^{\hspace{-0.5pt}\ast} f \rVert_{\rho}$, which is to say that the pullback is a contraction.

It now remains to introduce the dual notion of the pullback, which we call the {\it pushforward of the measurement} $M$.  To this, let us remark that the quotient norm on $S_{\rho}(\mathcal{H})$ admits a unique inner product $\langle A, B \rangle_{\!\rho} := \langle \{A, B\}\rangle_{\!\rho}/2$ compatible with the said norm in the sense $\lVert A \rVert_{\rho}^{2} = \langle A, A \rangle_{\!\rho}$, where we have used the anti-commutator $\{A,B\} := AB + BA$.  In the same manner, one may readily confirm that the inner product $\langle f, g \rangle_{p} := \langle fg \rangle_{p}$ defined on $R_{p}(\Omega)$ satisfies $\lVert f \rVert_{p}^{2} = \langle f, f \rangle_{p}$.  We then introduce the pushforward
\begin{equation}\label{def:pushforward}
M_{\!\rho\ast}^{\phantom{\hspace{-0.5pt}*}} : S_{\rho}(\mathcal{H}) \to R_{M\!\rho}(\Omega)
\end{equation}
as the adjoint of the pullback \eqref{def:pullback} with respect to the inner products.  
Again, in concrete terms, this implies that, given (the equivalence class of) a self-adjoint operator $A \in S_{\rho}(\mathcal{H})$, we have (that of) a corresponding real function $M_{\!\rho\ast}^{\phantom{\hspace{-0.5pt}*}} A \in R_{M\!\rho}(\Omega)$.  An important point to note is that the pushforward is defined as the unique map characterized by the relation
\begin{equation}\label{char:pf_and_pb_quantum-measurement}
\langle A, M_{\!\rho}^{\hspace{-0.5pt}\ast}f\rangle_{\!\rho}^{\phantom{\ast}}
	= \langle M_{\!\rho\ast}^{\phantom{\hspace{-0.5pt}*}} A, f\rangle_{\!M\!\rho}^{\phantom{\ast}}
\end{equation}
valid for all self-adjoint operators $A$ and real functions $f$.
Note also that the expectation value of an observable and that of its pushforward coincide $\langle A \rangle_{\!\rho} = \langle M_{\!\rho\ast}^{\phantom{\hspace{-0.5pt}*}} A \rangle_{\!M\!\rho}$, as can be readily confirmed by choosing the constant function $f = 1$ in \eqref{char:pf_and_pb_quantum-measurement}.  Since the pullback is a contraction, its adjoint, {\it i.e.}, the pushforward, is also a contraction $\lVert A \rVert_{\rho} \geq \lVert M_{\!\rho\ast}^{\phantom{\hspace{-0.5pt}*}} A \rVert_{M\!\rho}$.

%%%%%
%%%%%%%%%%
\prlsection{Error of Quantum Measurement\label{sec:error}}
%%%%%%%%%%
%%%%%
Armed with our geometric framework, let us introduce our definition of the (quantum) error by the amount of contraction induced by the pushforward of the measurement $M$,
\begin{equation}\label{def:error_quantum-measurement}
\varepsilon_{\rho}(A;M) := \sqrt{\lVert A \rVert_{\rho}^{2} - \lVert M_{\!\rho\ast}^{\phantom{\hspace{-0.5pt}*}}A \rVert_{M\!\rho}^{2}}
\end{equation}
for the observable $A$ and the quantum state $\rho$, both chosen independently from $M$.  Non-negativity $\varepsilon_{\rho}(A;M) \geq 0$ of the error follows from the contractivity of the pushforward.
It is also straightforward to check the homogeneity $\varepsilon_{\rho}(tA;M) = \lvert t \rvert\, \varepsilon_{\rho}(A;M)$, $\forall t \in \mathbb{R}$ and the subadditivity $\varepsilon_{\rho}(A;M) + \varepsilon_{\rho}(B;M) \geq \varepsilon_{\rho}(A + B;M)$ of the error.  In other words, the error furnishes a seminorm on the local space of quantum observables.

Our error also admits an operational interpretation as the minimal cost of the local reconstruction of quantum observables through quantum measurements.  To see this, let us consider the problem of reconstructing the observable $A$ locally by choosing an estimator function $f$ properly under the measurement $M$.  This may be implemented by using the pullback \eqref{def:pullback} of the measurement, which is a map that creates an observable $M_{\!\rho}^{\hspace{-0.5pt}\ast} f$ out of the function $f$ we choose.  We then introduce the error with respect to the estimator $f$ ({\it abbr.} $f$-error)
\begin{equation}
\varepsilon_{\rho}(A;M,f) := \sqrt{ \lVert A - M_{\!\rho}^{\hspace{-0.5pt}\ast}f \rVert_{\rho}^{2} + \Bigl( \lVert f \rVert_{M\!\rho}^{2} - \lVert M_{\!\rho}^{\hspace{-0.5pt}\ast}f \rVert_{\rho}^{2} \Bigr) }
\end{equation}
as a gauge of the precision of the reconstruction.  Here, the first term $\lVert A - M_{\!\rho}^{\hspace{-0.5pt}*}f \rVert_{\rho}$ of the gauge provides an evaluation of the algebraic deviation between the target and the reconstructed observables, while the second term $\lVert f \rVert_{M\!\rho}^{2} - \lVert M_{\!\rho}^{\hspace{-0.5pt}*}f \rVert_{\rho}^{2} = \sigma_{M\!\rho} (f)^{2} - \sigma_{\!\rho} (M_{\!\rho}^{\hspace{-0.5pt}*}f)^{2} \geq 0$ captures the increased cost in the reconstruction itself, which may have arisen from the suboptimal choice of the measurement $M$ or the estimator $f$ ({\it cf.}\ discussions below \eqref{ineq:contractivity}).
One then readily verifies by simple computation utilizing \eqref{char:pf_and_pb_quantum-measurement} that the square of the $f$-error admits the decomposition
\begin{equation}\label{eq:decomposition_error}
\varepsilon_{\rho}(A;M,f)^{2} = \varepsilon_{\rho}(A;M)^{2} + \lVert M_{\!\rho\ast}^{\phantom{\hspace{-0.5pt}*}}A - f \rVert_{M\!\rho}^{2}
\end{equation}
into the squared sum of the quantum and estimation errors.  
At this point, one inadvertently finds that the optimal estimator that minimizes the $f$-error is in fact given by $f = M_{\!\rho\ast}^{\phantom{\hspace{-0.5pt}*}}A$, which is the pushforward of the observable $A$ by the measurement $M$.  This provides an operational characterization of the quantum error
\begin{equation}\label{char:error}
\varepsilon_{\rho}(A;M)
	= \min_{f} \varepsilon_{\rho}(A;M,f)
\end{equation}
as the minimum over the $f$-errors, along with the interpretation of the pushforward as the locally optimal estimator.

%%%%%
%%%%%%%%%%
\prlsection{Uncertainty Relation for Error\label{sec:uncertainty_relation}}
%%%%%%%%%%
%%%%%
We are now ready to introduce our uncertainty relation.  In fact, one finds several inequalities that mark the trade-off relation between our measurement errors of incompatible quantum observables, 
but for the purpose of this Letter, we shall content ourselves with the simplest among them.
Let $A$ and $B$ be quantum observables, and $\rho$ be a quantum state on $\mathcal{H}$.  Then, for any quantum measurement $M: Z(\mathcal{H}) \to W(\Omega)$, the inequality
\begin{equation}\label{ineq:uncertainty_relation_error}
\varepsilon_{\rho}(A;M)\, \varepsilon_{\rho}(B;M)
	\geq \sqrt{\mathcal{R}^2 + \mathcal{I}^2}
\end{equation}
holds, where 
\begin{equation}\label{comp:urel_real}
\mathcal{R} := \left\langle \frac{\{A,B\}}{2} \right\rangle_{\!\!\!\rho} - \left\langle M_{\!\rho\ast}^{\phantom{\hspace{-0.5pt}*}}A, M_{\!\rho\ast}^{\phantom{\hspace{-0.5pt}*}}B\right\rangle_{\!M\!\rho}
\end{equation}
and
\begin{equation}\label{comp:urel_imaginary}
\mathcal{I} := \left\langle \frac{[A,B]}{2i} \right\rangle_{\!\!\!\rho}  - \left\langle \frac{[M_{\!\rho}^{\hspace{-0.5pt}\ast}M_{\!\rho\ast}^{\phantom{\hspace{-0.5pt}*}}A,B]}{2i} \right\rangle_{\!\!\!\rho} - \left\langle \frac{[A,M_{\!\rho}^{\hspace{-0.5pt}\ast}M_{\!\rho\ast}^{\phantom{\hspace{-0.5pt}*}}B]}{2i} \right\rangle_{\!\!\!\rho}
\end{equation}
with the commutator $[A,B] := AB -BA$.  

The proof of the inequality is actually quite simple: it is just a direct corollary of the Cauchy-Schwarz inequality.  A quick way to see this is to first introduce the semi-inner product
\begin{equation}\label{def:aux_semi-inner_product}
\langle (X, f), (Y, g) \rangle := \langle X^{\dagger}Y \rangle_{\!\rho} + \langle f^{\dagger} g \rangle_{\!M\!\rho} - \langle M^{\prime}f^{\dagger} M^{\prime}g \rangle_{\!\rho}
\end{equation}
defined on the product of the space of Hilbert space operators and that of complex functions, as well as the seminorm $p(X,f) := \sqrt{\langle (X, f), (X, f) \rangle}$ that it induces.  Noticing the equality
\begin{equation}\label{eq:aux_error_seminorm}
\varepsilon_{\rho}(A;M) = p(X_{A}, f_{A}),
\end{equation}
where $X_{A} := A - M_{\!\rho}^{\hspace{-0.5pt}\ast}M_{\!\rho\ast}^{\phantom{\hspace{-0.5pt}*}}A$ and $f_{A} := M_{\!\rho\ast}^{\phantom{\hspace{-0.5pt}*}}A$, we find that the Cauchy-Schwartz inequality for the product of $p(X_{A}, f_{A})$ and $p(X_{B}, f_{B})$ becomes
\begin{equation}\label{eq:csineq}
\varepsilon_{\rho}(A;M)\, \varepsilon_{\rho}(B;M)
	\geq \lvert \langle (X_{A}, f_{A}), (X_{B}, f_{B}) \rangle \rvert.
\end{equation}
The semi-inner product appearing in the R.H.S. of \eqref{eq:csineq} is a complex number $\langle (X_{A}, f_{A}), (X_{B}, f_{B}) \rangle = \mathcal{R} + i\,\mathcal{I}$ whose real part
$\mathcal{R}$ is given by \eqref{comp:urel_real} while the imaginary part $\mathcal{I}$ by \eqref{comp:urel_imaginary}.  
This completes our proof of the inequality \eqref{ineq:uncertainty_relation_error}.
 
From a geometric point of view, the real part $\mathcal{R}$ in \eqref{comp:urel_real} represents the diminished metric on the bundle of quantum observables inevitably caused by the measurement $M$.  We also note that the invariance of the expectation value under the pushforward allows us to interpret $\mathcal{R} = \mathrm{Cov}_{\rho}(A,B) - \mathrm{Cov}_{M\!\rho}(M_{\!\rho\ast}A,M_{\!\rho\ast}B)$ as the difference between the covariances.  In fact, as one may demonstrate through parallel arguments, this term is also found to be shared with classical measurements, which suggests that the real part $\mathcal{R}$ is not necessarily of quantum origin.  On the other hand, the imaginary part $\mathcal{I}$ in \eqref{comp:urel_imaginary}, which consists of three commutators and gives an additional contribution to the lower bound, marks the essence of quantum measurements.  In this regard, the reduced simpler form $\varepsilon_{\rho}(A;M)\, \varepsilon_{\rho}(B;M) \geq \lvert \mathcal{I} \rvert$ obtained from \eqref{ineq:uncertainty_relation_error} should be sufficient to express its distinctive characteristics.

%%%%%
%%%%%%%%%%
\prlsection{The Uncertainty Principle}
%%%%%%%%%%
%%%%%
Our uncertainty relation \eqref{ineq:uncertainty_relation_error} implies a potential violation of the non-commutativity bound $\lvert \langle [A, B] \rangle_{\!\rho} / 2i \rvert$ for certain choices of quantum measurements.  It is to be emphasized, however, that even though the product of the errors may overcome the non-commutativity bound quantitatively, the philosophy of the uncertainty principle remains valid: simultaneous errorless measurement of non-commutative observables is impossible when $\langle [A, B] \rangle_{\!\rho} \neq 0$.  We shall now argue why this is the case.

For this, we need to discuss the situation where the measurement $M$ becomes free from the error.  
We say that a quantum measurement $M$ is capable of an errorless measurement of $A$ over $\rho$, if the error $\varepsilon_{\rho}(A;M)$ vanishes.  Several characterizations of the errorless measurement are possible, and here we note the equivalence of the following three conditions:
\begin{enumerate}[label=\rm{(\alph*)}]
\item $\varepsilon_{\rho}(A;M) = 0$,\label{fact:errorless-measurement_1}
\item $A = M_{\!\rho}^{\hspace{-0.5pt}\ast}M_{\!\rho\hspace{0.5pt}*}^{\phantom{*}} A$,\label{fact:errorless-measurement_2}
\item $\lVert A \rVert_{\rho} = \lVert M_{\!\rho\hspace{0.5pt}*}^{\phantom{*}} A \rVert_{M\!\rho} = \lVert M_{\!\rho}^{\hspace{-0.5pt}\ast}M_{\!\rho\hspace{0.5pt}*}^{\phantom{*}} A \rVert_{\rho}$.\label{fact:errorless-measurement_3}
\end{enumerate}
In fact, $\ref{fact:errorless-measurement_3} \implies \ref{fact:errorless-measurement_1}$ is trivial by definition, $\ref{fact:errorless-measurement_1} \implies \ref{fact:errorless-measurement_2}$ is an immediate consequence of \eqref{eq:decomposition_error} with the substitution $f = M_{\!\rho\ast}^{\phantom{\hspace{-0.5pt}*}}A$, and finally $\ref{fact:errorless-measurement_2} \implies \ref{fact:errorless-measurement_3}$ is due to the contractivity $\lVert A \rVert_{\rho} \geq \lVert M_{\!\rho\hspace{0.5pt}*}^{\phantom{*}} A \rVert_{M\!\rho} \geq \lVert M_{\!\rho}^{\hspace{-0.5pt}\ast}M_{\!\rho\hspace{0.5pt}*}^{\phantom{*}} A \rVert_{\rho} = \lVert A \rVert_{\rho}$ of both the pullback and the pushforward.

An immediate corollary of this is that, for a non-commuting pair of observables $A$ and $B$, there is no quantum measurement that is capable of measuring both observables errorlessly, $\varepsilon_{\rho}(A;M) = 0$ and $\varepsilon_{\rho}(B;M) = 0$, over $\rho$ for which the non-commutativity term $\langle [A, B] \rangle_{\!\rho}$ is non-vanishing.  Indeed, if there were such a measurement, our uncertainty relation \eqref{ineq:uncertainty_relation_error} combined with the equivalence $\ref{fact:errorless-measurement_1} \iff \ref{fact:errorless-measurement_2}$ would lead to a contradiction $0 \geq \sqrt{\lvert 0 \rvert^{2} + \lvert \langle [A, B] \rangle_{\!\rho}/2i \rvert^{2}}$.  Another way to put it is that, for non-trivial ({\it i.e.}, $\mathrm{dim}(\mathcal{H}) \geq 2$) quantum systems, there exists no quantum measurement that is capable of errorlessly measuring every quantum observable over every quantum state.
Note that our formulation does not necessarily prohibit one of the errors from vanishing.  This is in contrast to other formulations including the KR inequality
that respect the non-commutativity bound, in which a stronger restriction holds so that neither of the terms may vanish.  

%%%%%
%%%%%%%%%%
\prlsection{Reference to Other Uncertainty Relations}
%%%%%%%%%%
%%%%%
Since our uncertainty relation is established on a very simple and general set of premises of quantum measurement, it is worthwhile to consider whether it can shed some light on other notable uncertainty relations mentioned in the Introduction.

In this respect, we first show that the KR inequality actually emerges as a trivial case of our relation.  We may call a quantum measurement $M$ trivial, or non-informative, when it is a constant map, {\it i.e.}, $M\!\rho = p_{0}$ for all $\rho \in Z(\mathcal{H})$ with some fixed $p_{0} \in W(\Omega)$.  In other words, trivial measurements are the least informative measurements one could possibly make on a quantum system.  It is fairly straightforward to confirm that the pushforward of an observable $A$ by any trivial measurement is the constant function $M_{\!\rho\ast}^{\phantom{\hspace{-0.5pt}*}}A = \langle A \rangle_{\!\rho}$ of the observable's expectation value.  Triviality of the measurement thus reduces our error to the standard deviation $\varepsilon_{\rho}(A;M) = \sigma_{\!\rho}(A)$, further bringing our overall uncertainty relation \eqref{ineq:uncertainty_relation_error} down to
\begin{equation}
\sigma_{\!\rho}(A)\, \sigma_{\!\rho}(B) \geq \sqrt{\mathcal{R}^2 + \mathcal{I}^2}
\end{equation}
with $\mathcal{R} = \langle \{A,B\} \rangle_{\!\rho}/2 - \langle A \rangle_{\!\rho}\langle B \rangle_{\!\rho}$ and $\mathcal{I} = \langle [A,B] \rangle_{\!\rho}/2i$.
%\begin{equation}
%\sigma_{\!\rho}(A)^2 \sigma_{\!\rho}(B)^2
%	\geq 
%	\left\lvert \left\langle \frac{\{A,B\}}{2} \right\rangle_{\!\!\!\rho} - \langle A \rangle_{\!\rho}\langle B \rangle_{\!\rho} \right\rvert^{2}
%	+ \left\lvert \left\langle \frac{[A,B]}{2i} \right\rangle_{\!\!\!\rho} \right\rvert^{2}.
%\end{equation}
This is known as the Schr{\"o}dinger inequality \cite{Schroedinger_1930}, from which the KR inequality follows immediately.
We thus have observed that, through the process of rendering the measurement into triviality, our inequality finds a seamless connection between the two different realms of uncertainty relations: one regarding measurement errors and the other regarding quantum indeterminacy expressed by standard deviations.

We next note that our framework naturally encompasses the indirect measurement scheme adopted by several alternative formulations, for every quantum measurement employing detector systems also  preserves the structure of probabilistic mixture.  Under such model, Ozawa proved \cite{Ozawa_2004_01} the inequality $\varepsilon(A) \varepsilon(B) \geq \lvert \left\langle [A,B]\right\rangle_{\!\rho} \rvert /2 - \varepsilon(A)\sigma(B) - \sigma(A)\varepsilon(B)$ for joint measurements of $A$ and $B$, where $\varepsilon(A)$ and $\varepsilon(B)$ are his errors for the respective observables and his $\sigma$ is the same as our $\sigma_{\!\rho}$.  In fact, our uncertainty relation, with suitable refinements to accommodate joint measurability, is found to reduce Ozawa's relation to one of its corollaries.  A simple way to explain this is to demonstrate that our relation is tighter than Ozawa's: one finds that Ozawa's error is never less than ours, and further reveals $\varepsilon(A) \varepsilon(B) \geq \varepsilon_{\rho}(A) \varepsilon_{\rho}(B) \geq \sqrt{\mathcal{R}^2 + \mathcal{I}^2} \geq \lvert \mathcal{I} \rvert \geq \lvert \left\langle [A,B]\right\rangle_{\!\rho} \rvert /2 - \varepsilon(A)\sigma(B) - \sigma(A)\varepsilon(B)$.  Here, the short forms $\varepsilon_{\rho}(A)$ and $\varepsilon_{\rho}(B)$ denote our errors regarding the respective observables, and $\mathcal{R}$, $\mathcal{I}$ are the terms respectively related to \eqref{comp:urel_real} and \eqref{comp:urel_imaginary} that marks the lower bound of the product of our errors under joint measurement.  As should be expected, AKG's relations, which is valid under additional unbiasedness condition assumed on top of the measurement model adopted by Ozawa, can also be framed as a corollary to ours.  Details on this topic will be reported in our subsequent papers.

\begin{acknowledgments}
The authors thank Prof.~Naomichi Hatano for fruitful discussions and insightful comments.
This work was supported by JSPS Grant-in-Aid for Scientific Research (KAKENHI), Grant Numbers JP18K13468 and JP18H03466.
\end{acknowledgments}

\bibliography{urel}

\end{document}